\documentclass[preprint,12pt]{elsarticle}

\usepackage{amssymb}
\usepackage{url}
\usepackage{booktabs} 

\usepackage{hyperref}
\usepackage{caption}
\usepackage{rotating}
\usepackage{graphicx}
\usepackage{multirow}
\usepackage{array}
\usepackage{amsmath}

\begin{document}

\begin{frontmatter}


\author[1]{Areeb Ahmed Mir}
\ead{areeb.se@must.edu.pk}
\author[1]{Muhammad Raees}
\ead{raees.se@must.edu.pk}
\author[1]{Afzal Ahmed}
\ead{afzal.se@must.edu.pk}

\affiliation[1]{organization={Mirpur University of Science and Technology},
            city={Mirpur},
            state={AJK},
            country={Pakistan}}

\title{Object Oriented-Based Metrics to Predict Fault Proneness in Software Design}

\begin{abstract}
In object-oriented software design, various metrics predict software systems' fault proneness. 
Fault predictions can considerably improve the quality of the development process and the software product. 
In this paper, we look at the relationship between object-oriented software metrics and their implications on fault proneness.
Such relationships can help determine metrics that help determine software faults.
Studies indicate that object-oriented metrics are indeed a good predictor of software fault proneness, however, there are some differences among existing work as to which metric is most apt for predicting software faults.
\end{abstract}


\end{frontmatter}


\section{Introduction}
Software fault prediction models are used when there is a need to predict faults in software that might arise, to preempt those faults, and to improve the quality of software. 
Various modeling techniques are used by studies with varying object-oriented metrics. 
However, the difference in modeling technique does not produce as much performance difference as the choice of metrics does when it comes to predicting the faults in software~\cite{1radjenovic2013software, raees2020study}.
For predicting object-oriented software faults, many object-oriented metrics have been proposed, however, they have limited use in practice despite applying to very complex software systems. 
Nevertheless, because of the variance in the performance of prediction due to different metrics, appropriate metrics for predicting faults are important to consider.
In this study, we look at extracting the relationship between object-oriented software metrics and the software fault proneness and identify which metrics can be successfully used to predict the software fault proneness. 
We examined some studies from the past and analyzed those to extract patterns for evaluating the fault predictor object-oriented metrics. 
We analyzed which metrics can be used for predicting faults and identified the most commonly used metrics that can accurately predict software fault proneness.

Danijel et al. \cite{1radjenovic2013software} carried out an extensive systematic literature review.
The objective of their review was to assess the primary studies that validate software metrics empirically for predicting software fault and to assess the metrics used in the selected studies according to various properties. 
They found out that Object-oriented metrics were used 49 \% of the time, which is nearly twice as much as compared to the traditional source code metrics used 27 \% and process metrics which is used 24 \%. 
Their study revealed that Chidamber and Kemerer’s (CK) object-oriented metrics were used most frequently. 
They found that there are substantial differences between fault prediction performance and the metrics used for prediction. 
They found object-oriented and process metrics were more successful in predicting faults compared to the traditional metrics.
We aimed at finding out the relationship between object-oriented metrics and software fault proneness to determine which of these metrics can be successfully used to predict the fault proneness of software.

\section{Research Dimension}
\label{section:3method}
To summarize the current research in this field of the prediction of faults in software using object-oriented metrics, we have performed a literature analysis. 
Our approach of carrying out the literature review was slightly like the one used by Danijel et al.~\cite{2huang2009multi, raees2024explainable}.
Our review was performed in three stages: planning, executing, and then reporting. 
We identified and reviewed existing systematic reviews on the topic. 
We defined our research questions, which compared with the existing research. 
We defined the search strategy and study selection process (inclusion and exclusion criteria). 
The data was then extracted and analyzed, and results were constructed and reported in this research paper.
In our analysis, we looked to answer the following research questions:

\begin{itemize}
    \item Research Question 1: What is the relationship between object-oriented software metrics and software fault proneness?
    \item Research Question 2: Which of these metrics can be successfully used to predict the software fault proneness?
\end{itemize}

With these questions, we studied our primary studies and extracted the relationship between various object-oriented metrics. 
During the search process, we retrieved an initial list of our existing studies. 
The following general search string was used: (object oriented OR object-oriented OR OO) AND (software) AND (predict OR predicting) AND (fault proneness OR fault-proneness). 

\section{Literature Analysis}
Arashdeep et al.~\cite{3kaur2009early} conducted a study to find defects early in the software life cycle using defect data of real-time software. 
The study aimed to find out whether metrics are available in the early life-cycle of software, i.e. the requirement metrics and the metric available in later stage of software life-cycle.
The combination of both can be used to identify fault-prone modules in object-oriented real-time software using clustering techniques. 
This approach was tested on three projects, and their data was collected from the NASA MDP (Metric Data Program) data repository. 
The projects chosen were CMI (spacecraft instrument program), JMI (real time ground system), and PCI (an earth orbiting satellite).
The software was divided into two categories: fault-prone and fault-free.
To identify whether the software was fault-free or fault-prone, a clustering algorithm was applied to the data. 
The evaluation measures used were as follows:
\begin{itemize}
    \item Probability of detecting (P D) = T P / (T P + F N)
    \item Probability of false-alarm (P F) = F P / (F P + T N)
\end{itemize}

ROC (Receiver Operating Characteristic) was plotted, which is a plot of P D vs. P F graph. 
This ROC curve’s shape is analyzed to determine the result. 
This study was fruitful, but it only used one algorithm for prediction. 
Peng et al.~\cite{2huang2009multi} conducted an empirical study in which they applied a multi-instance (MI) prediction model for predicting the fault-proneness in a software system. 
The study aimed to investigate the effects of multi instance learning in object oriented software’s quality. 
Five data sets were obtained from an industrial optical communication project. 
According to the MI definition, each class was regarded as an instance, and each class that had an inheritance relation from some other class was regarded as a bag. 
For evaluation purposes, each class was represented by a vector~\cite{dar2016characterizations}.

The task of this model was to label each untested class as either faulty (negative) or non-faulty (positive). 
The function of the MI kernel was to compute the inner product of the two classes (represented using vectors). 
This kernel was used to predict the faulty or non-faulty classes. 
The results showed that the MI kernel method performed better than the MI learning algorithms and that the class hierarchy metric is a good measure for predicting fault-proneness in software. 
The class hierarchy metric combined with the kernel method proved to be beneficial for predicting faults even in smaller systems that are in progress.
Ana~\cite{4camargo2010exploratory} conducted an exploratory study of a UML metric for fault prediction of object-oriented software. 
The study used the UML RFC (response for class) metric to predict faults in the software. 
RFC is defined as a collection of all the functions that can be called in response to a message call to object of a class. 
The RFC in this study was determined from UML collaboration diagrams. 
The study used three small-sized software programs developed by students of the researcher for their evaluation. 
They measured the RFC metric from UML diagrams as well as directly from code and compared them. 
To allow for the proposed method to be used for other research as well, the author proposed and normalized the difference between UML and code measure using Linear Scaling to Unit Variance. 
RFC measure was the independent variable, and fault in class was the dependent variable having two states: MF (most faulty) and LF (least faulty).
After setting these variables, the author then performed Logistic Regression to build the prediction model. 
Once the model was built, it was then applied to various other open source projects, whose data is freely available online, to determine the correctness of the model. 
The results showed that the UML RFC metric is as good at predicting faults in software as the code RFC metric is at predicting faulty code. 
The study was pretty much thorough and even evaluated itself by validating the prediction model by applying it on different data sets.

Marshima et al.~\cite{5rosli2011design} presented a research in which they proposed a design of an application that uses a computer-aided approach to predict the fault proneness of software when object-oriented metrics are input to it, which it will use along with metric values from an open-source system. 
The proposed application will use a genetic algorithm to classify a software module as either faulty or non-faulty. 
The proposed evolutionary algorithm was implemented in java and then applied on defect data set of JMonkeyEngine which is an open source java game engine After application of algorithm on the data set the results were modeled using bar and pie chart which showed that 77 \% of class is non-faulty and 23 \% of classes are faulty. 
This research showed that both object-oriented and count metrics can be used for the prediction of faults in software, however, the study was very silent as to which object-oriented metrics it employed, which is a drawback of this study and needs to be answered.
Santosh et al. ~\cite{6rathore2012investigating} conducted an empirical study to investigate the relationship between class design level object-oriented metrics and the fault proneness of the software. 
The study focused on coupling, cohesion, complexity, inheritance, and size metrics in predicting the faults in software. The relationship between these metrics and the fault-proneness was tested both individually and using a combination of these metrics. 
To investigate the effect of metrics, the metrics were first categorized according to the design attributes, and the first set of tests was carried out with metrics considered in isolation. 
The defect data used was acquired from the Promise Data Repository, and the prediction models were applied to the acquired data. 
For the isolated metrics evaluation, first, Univariate Logistic Regression was performed, followed by finding correlation using Spearman’s correlation, and then AUC (Area under ROC curve) analysis was performed to find which metric is more significant in finding faults, in isolation.
Next the metrics were considered in different combinations and were evaluated for prediction of faults in software using multivariate prediction model. 
Both prediction models used four machine learning techniques namely: Navie Baise, Logistic Regression, Random Forest and IBK classification for prediction of faults.
However, the combination of these metrics was evaluated not for figuring out which combination is the best predictor but rather to measure the accuracy of metrics prediction. 
The results showed that coupling, complexity, and size, to some extent, are the main attributes that are significant predictors of fault, not just in terms of precision but also in terms of accuracy. 
The systematic combination of these metrics yields a higher accuracy for predicting the faults in software.

Santosh et al.~\cite{7rathore2012validating} conducted another study like the one conducted in ~\cite{6rathore2012investigating} with a slightly different approach; the study aimed to find the effectiveness of object-oriented metrics in predicting faults in software over multiple releases. For the empirical analysis, the data was once again acquired from the Promise Data Repository, and the projects chosen were Camel, Xalan, Xerces, Ivy, and Velocity. 
The criterion for fault proneness for multiple releases was defined as the occurrence of a fault in the newer release when moving from one release to another. 
Like the study conducted in ~\cite{6rathore2012investigating}, the authors performed Univariate Logistic Regression as well as Multivariate Logistic Regression as well as cross-correlation to find out which metrics are correlated to each other.
Like the study in ~\cite{6rathore2012investigating}, this study also used the same four machine learning techniques. 
The results showed that moving from one release to the other the metrics which were significant in one release changed in some cases for the next release. 
However, the authors chose only those metrics that were significant in all releases. The study showed that it is possible to select a small subset of metrics that can predict faults with high accuracy and precision. 
In this case, those metrics were WMC, CBO, RFC, LOC, CA, and CE. However, if this study had been conducted on a larger scale, i.e. using more software’s defect data, it could have come up with a more generalized subset of metrics which are true predictors of faults in software.

Arwin ~\cite{8halim2013predict} conducted a study to predict fault-prone classes using the complexity of the UML Class diagram. 
The author built the prediction model using two classification algorithms – k nearest neighbors (kNN) and Naïve Bayesian. 
Both the models were validated using 10-fold cross-validation, and the performance of these prediction models was evaluated using Receiver Operating Characteristic (ROC) analysis. The data needed for conducting this study was available online – data of four versions of open source software JEdit was used for this study.
However only source code was available and not the UML diagram so the author used UModelTM software to reverse engineer the UML class diagrams from the java source code. 
The metrics required for this study class complexity (CC), relationship complexity (RC) and the no. of class relationship (NC) were computed using automated tool AHComplex. 
The metrics values were then fed into the fault-prediction model. Each dataset was divided into 10 subsamples for 10-fold cross-validation by using each data subsample for data testing and training, iteratively. 
The results indicated that the prediction model based on design complexity can predict fault-prone classes using the Naïve Bayesian algorithm with a 77.3 \% success rate and with 77.8 \% accuracy using k-nearest neighbors. 
Overall, the study showed that the complexity of the UML class diagram can be used to predict fault-prone classes successfully no matter which algorithm is used. However, this study had one disadvantage, and that is it applied the model on the same data through which they were derived, and if these models are applied on some other datasets as well, that will be better for the validation of such model.

Raed ~\cite{9shatnawi2014empirical} conducted an empirical study on open source software using CK metrics to predict fault proneness in classes of software. 
The study is unique from others in two ways: it was carried out on not only one version of each of the four open source software that it used (Ant, Camel, JEdit and Xerces) rather it was carried out on four versions of each software to bring into account the history of faults of classes in previous release of software.
Secondly the study didn’t just use a binary variable (i.e. fault/non-faulty) rather it categorized classes into four categories: none, low risk, medium risk and high risk of fault to improve the predictive capability of fault prediction models. 
The study used 7 machine learning algorithms (i.e. Bayes net, neural network, support vector machines, nearest neighbors, C4.5 decision trees, random forest trees and CART decision trees) and compared the results achieved by each to determine which prediction model gives better results. 
A total of six metrics was used for prediction of faults: CBO (coupling between objects), DIT (depth of inheritance tree), NOC (number of child classes), LCOM (lack of cohesion metrics), RFC (response for class) and WMC (weighted methods complexity).
The fault data needed for evaluation was taken from Promise data repository and the fault data was extracted from the repository using two tools Buginfo and Ckjm. 
The threshold value was set to 0.5, and probability greater than the threshold classified the class as faulty. The results showed that the multi-categorization of faulty classes is better than the binary variable (faulty/non-faulty) for the same metrics. 
The study also indicated that LCOM (lack of cohesion metrics) is not such a good predictor of faults in software as it gave same values for classes with different cohesion values. 
The study was very thorough and detailed and the fact that it was repeated for various versions of same software gives it an upper hand and produced more realistic conclusion.

Jeenam et al. ~\cite{10techobject} reviewed various object-oriented design metrics that can predict the fault proneness of software. 
The study reviewed some of the studies carried out by other researchers and reported the metrics used by those researchers to predict the fault proneness of classes and presented the metrics that are useful as predictors. 
The various metrics presented by this study are: WMC (weighted methods in each class) i.e. weighted sum of all methods of a class, CBO (coupling b/w objects) i.e. count of coupling between the classes, DIT (depth of inheritance tree) i.e. the length of the longest leaf from root node in inheritance hierarchy, NOC (number of children) i.e. count of children that a class has, RFC (response for class) i.e. collection of all the functions that can be invoked in response to a message call to object of a class etc.
One study showed that CBO, DIT, WMC, and NOC were highly correlated with fault-proneness, while one study showed that CBO was the best predictor, followed by LOC metric. 
Some studies’ results complemented others’, while some appeared to contradict others. 
This contradiction was mostly due to the choice of dataset used by various authors and not due to different prediction models employed.

\section{Summary}
In this paper, we reviewed various studies presenting the relationship between object-oriented metrics and the fault proneness of object-oriented software. 
The studies reviewed show the diversity in the multitude of metrics chosen to predict the fault proneness of the software. 
Some studies did, however, show common metrics being used for evaluation. 
Such studies did indicate corroborative evidence that coupling, complexity and inheritance metrics are the most significant predictors of faults in software. 
However, contradicting evidence was also found conclude that coupling, complexity and inheritance metrics are de facto the absolute predictors of faults for all software, rather they are the predictors of faults for most software.

 \bibliographystyle{elsarticle-num} 
 \bibliography{cas-refs}





\end{document}